\documentclass[prl,twocolumn,superscriptaddress,showpacs,letter]{revtex4}
\usepackage{graphics}
\begin{document}
\title{The r\^{o}le of electronic correlations on the the phonon modes of MnO and NiO}
\author{E. M. L Chung}
\affiliation{Department of Physics, University of Warwick, Coventry, CV4 7AL, UK}
\author{D. M$^{c}$K. Paul}
\affiliation{Department of Physics, University of Warwick, Coventry, CV4 7AL, UK}
\author{G. Balakrishnan}
\affiliation{Department of Physics, University of Warwick, Coventry, CV4 7AL, UK}
\author{M. R. Lees}
\affiliation{Department of Physics, University of Warwick, Coventry, CV4 7AL, UK}
\author{A. Ivanov}
\affiliation{Institut Laue Langevin, BP 156, 38042, Grenoble Cedex 9, France}
\author{M. Yethiraj}
\affiliation{Oak Ridge National Laboratory, Tennessee, 37831, United States}
\date{\today}

\begin{abstract}
The possibility of magnetic-order induced phonon anisotropy in single
crystals of MnO and NiO is investigated using inelastic neutron
scattering. Below $\mathrm{T_{N}}$ both compounds exhibit a splitting
in their transverse optical phonon spectra of approximately 10\%.
This behavior illustrates that, contrary to general assumption, the
dynamic properties of MnO and NiO are substantially non-cubic.
\vspace{8mm}
\end{abstract}

\pacs{63.20.-e, 75.30.Gw, 78.70.Nx}

\maketitle

The failure of \emph{ab initio} approximations to correctly incorporate many-body effects 
such as electron exchange and correlation is an issue at the core of contemporary solid-state 
research. Although these effects are important in almost all solids, they are essential 
for a proper description of co-operative phenomena such as antiferromagnetism, charge-ordering, 
superconductivity and colossal magnetoresistance. 

Despite the status of MnO and NiO as 
benchmark materials for the study of correlated electron systems and frequent use in first 
principles electronic structure investigations, many aspects of the physics of the 3$d$ 
transition metal monoxides requires better theoretical explanation. For example, the basic 
physical properties predicted for MnO and NiO (e.g. band-gaps, 
distortion angles and phonon spectra) differ radically depending upon which techniques 
have been employed  \cite{towleretal1994,massiddaetal1999}.

In a recent publication comparing different \emph{ab initio} and model band-structure models on 
MnO (including local spin-density approximation (LSDA) and `LSDA + model' calculations), 
Massidda \emph{et al.} \cite{massiddaetal1999} suggest that although the electronic density in 
MnO is approximately cubic, the lowering of symmetry associated with antiferromagnetic ordering 
can induce an electronic response that is significantly non-cubic. This results in the 
dynamical properties of the system, such as transverse optical (TO) phonons, 
exhibiting substantial ``\emph{magnetic-order induced anisotropy}". 

\begin{table}[b]
\begin{ruledtabular}
\begin{tabular}{llll}
METHOD & TO Mode(s) 	& \\
\hline
LCAO UHF  	& 38.95 meV & Towler \emph{et al.} \cite{towleretal1994}\\
\hline
(LAPW) LSDA   	& 12.46 meV & Massidda \emph{et al.} \cite{massiddaetal1999}\\
\hline
LSDA + model 	& 28.6 meV \& & \\
 & 33.9 meV & Massidda \emph{et al.}\cite{massiddaetal1999}\\
\hline
Inelastic neutron &   32.5 meV & Haywood \emph{et al.} \cite{HaywoodandCollins1969} \\
scattering (300 K) & 32.2 meV & Wagner \emph{et al.} \cite{Wagneretal1976}\\ 
      &   33.2 $\mathrm{\pm}$ 0.2 meV &  \textbf{This work}         \\
\hline
Inelastic neutron & 33.3 $\mathrm{\pm}$ 0.2 meV \& &\\
scattering (4.3 K)& 36.4 $\mathrm{\pm}$ 0.3 meV &  \textbf{This work}  \\
\hline
Optical absorption & &  \\
( below $\mathrm{T_{N}}$) & 32.54 meV &Yokogawa \emph{et al.} \cite{Yokogawaetal1977} \\
\end{tabular} 
\end{ruledtabular}
\caption{Comparison of theoretical predictions and experimentally observed values for the energies 
of transverse optical phonons in MnO. Our results are in good agreement with previous 
experimental work, with the exception of the splitting, which is qualitatively consistent 
with `LSDA + model' calculations by Massidda \emph{et al.}\cite{massiddaetal1999} }
\label{tab1}
\end{table}

One of the predicted 
features from such calculations is that the zone center (ZC) optical 
phonon modes should split depending on their polarization. A higher energy occurs for 
a mode polarized along the [111] direction and a lower energy for degenerate modes polarized 
in the orthogonal ferromagnetic plane, see Table \ref{tab1}. Massidda \emph{et al.} \cite{massiddaetal1999} 
find that this effect arises \emph{solely due to the magnetic ordering}, and would 
even occur in the absence of any rhombohedral distortion. By comparing results from four 
different approximation schemes, lower and upper bounds of 3-10\% for the magnitude of the splitting 
are estimated. Variations in the splitting due to the distortion are calculated to be less than 1\%.

\begin{figure}[t]
\begin{center}
\rotatebox{90}{ }\resizebox{8cm}{!}{\includegraphics{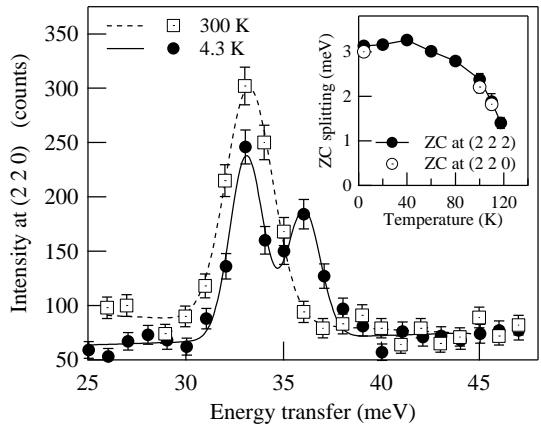}}
\caption{Zone-center optic phonon modes measured at 4.3 K (full line, filled markers) and 300 K 
(dashed line, open markers). (For both scans mon=1500). The measurements shown are representative 
of scans obtained at \textbf{Q}=(220), \textbf{Q}=(004), \textbf{Q}=(222) and \textbf{Q}=(113). 
A higher energy mode (possessing a third of the intensity of the original peak) is 
created at low temperature due to the magnetic ordering. The inset graph shows the temperature 
dependence of the magnitude of the MnO zone-center TO phonon 
splitting. The splitting is clearly associated with the onset of antiferromagnetic order at 
T$_{N}$=118 K. }
\label{zcsplitting}
\end{center}
\end{figure}

Although similar theoretical calculations have not yet been performed for NiO, the results 
of measurements of NiO TO phonons are predicted to be qualitatively similar to those in MnO. 
Indeed, it has been suggested that the splitting in NiO may be even greater in absolute magnitude 
than in MnO due to the larger magnetic super-exchange \cite{Massidda2001}. 

MnO and NiO are classic examples of type-II antiferromagnets possessing the cubic rock-salt structure. 
Below $\mathrm{T_{N}}$, exchange-striction causes contraction of the unit cell along the [111] direction 
perpendicular to the ferromagnetic planes \cite{GreenwaldandSmart1950,Roth1958}. This results in a small 
deviation from the perfect cube of 0.62$\mathrm{^{o}}$ for MnO and 0.1$\mathrm{^{o}}$ for NiO. In MnO 
the magnitude and temperature dependence of the rhombohedral distortion has been well 
characterized \cite{Morosin1970}. Since the distortion is considered weak, investigations of the 
dynamic properties of MnO and NiO are frequently carried out under the assumption 
of perfect cubic symmetry.

To examine the possibility of anisotropy in the optical phonon spectra of MnO (and NiO), 
(and distinguish these effects from any changes due to the small rhombohedral distortion), 
TO phonons in single crystals of MnO and NiO were measured using inelastic neutron scattering. 
MnO and NiO crystals used in these experiments were grown using the floating-zone method 
at a rate of 10 mm/h in 1 atm of Argon. Both crystals consisted of small cylindrical pillars 
approximately 0.5 cm in diameter and 2 cm long. 

The optic phonons, magnon modes and sublattice magnetization of our 
MnO crystal were studied using the IN8 triple axis spectrometer located at the high-flux 
reactor at the Institut Laue Langevin, (ILL), Grenoble. The Cu(111) reflection was used as 
monochromator in combination with a pyrolytic graphite (PG) 
filter, PG(002) analyzer and Soller collimation. Measurements were made using constant-\emph{Q} scans, with a fixed final 
energy of $\mathrm{E_{f}}$ = 14.69 meV, ($\mathrm{k_{f}}$ = 2.66 \AA). The crystal was oriented 
with the [110] direction vertical. 

Operating the spectrometer in diffraction mode, measurements of the temperature evolution of 
the magnetic ($\frac{3}{2}$ $\frac{3}{2}$ $\frac{3}{2}$) reflection were obtained. The 
observed value of $\mathrm{T_{N}}$= 118 K, measured during slow cooling at a rate of 
1$\mathrm{^o}$ every 3 mins, was in excellent agreement with previous 
reports \cite{Blochetal1974,BoireandCollins1977,Woodfieldetal1999}.  

Compression along equivalent [111] directions in the bulk crystal creates four possible 
orientations of twinning, or `T-domain', each with a distinct 3\emph{D} orientation of 
magnetic Brillouin zone. The scattering plane of the sample is then comprised of a 
superposition of slices through each of the magnetic Brillouin zones \cite{Kohgietal1974,Bonfanteetal1972}. 
The T-domains in our MnO sample were characterized using spin wave dispersion measurements 
obtained at 4.3 K. Proceeding by continuity from unambiguously defined reciprocal lattice positions, 
the full magnon dispersion curves of our multi-T-domain MnO crystal along the [110], [001] and [111] 
directions were determined. From these dispersion curves it was possible to deduce that the crystal 
was comprised of two approximately equally populated twinned domains.

\begin{figure}[b]
\begin{center}
\resizebox{70mm}{!}{\includegraphics{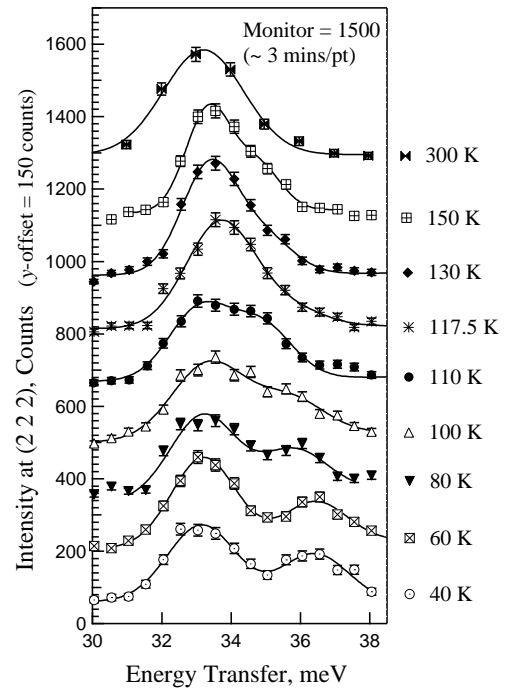}}
\caption{Waterfall plot showing the temperature dependence of the MnO TO 
phonon spectra at \textbf{Q}=(222). (Scans have been offset by an arbitrary y-value of 150 counts.)}
\label{fulltdep}
\end{center}
\end{figure}

Measurements of transverse optical (TO) phonon modes were obtained along several directions 
in the crystal ([001], [111] and [110]) and around various Brillouin zone (BZ) centers 
(\textbf{Q}=(004), \textbf{Q}=(113), \textbf{Q}=(222) and \textbf{Q}=(220)). At 4.3 K, splitting of 
the modes was consistently observed in all of the measured BZ's. Representative zone center 
scans obtained at 4.3 K and 300 K are shown in figure 1. 

The energy of the 300 K ZC excitation occurs at 33.2 $\mathrm{\pm}$ 0.2 meV with a 
full-width half-maximum of 2.87 $\mathrm{\pm}$ 0.5 meV. In the antiferromagnetically 
ordered phase at 4.3 K two modes were consistently observed at energies of 
E = 33.3 $\mathrm{\pm}$ 0.2 meV and E = 36.4 $\mathrm{\pm}$ 0.3 meV. The zone center 
splitting at 4.3 K is approximately 3.1 meV, which corresponds to 9.3\% of the energy 
of the original (degenerate) mode above $\mathrm{T_{N}}$. 

The uncertainty in reduced wave vector, \textbf{q}, arising from the rhombohedral 
distortion was estimated for the measured Brillouin zones to range between \emph{d}\textbf{q}= 0.02 
and \emph{d}\textbf{q}= 0.04 r.l.u. At the ZC position where the dispersion of the TO modes is virtually 
flat the effects of an uncertainty of this magnitude on the energy spectra would be negligible. 
We therefore conclude that the splitting is a real effect which can not be attributed to the 
small static structural distortion.

The low temperature data were fitted to a pair of Gaussian peaks of identical width. 
Peaks at each of the zone-centers possessed widths of 2.43 $\mathrm{\pm}$ 0.6 meV. 
The ratio of the integrated intensities of the two peaks was approximately 2:1 in 
favor of the lower mode at each of the measured zone-centers. The total integrated 
intensity of the two (split) peaks and the initial single peak were approximately 
conserved on passing through T$_{N}$. This behavior is consistent 
with the creation of a new higher energy mode arising from excitations polarized 
along the [111] direction \cite{massiddaetal1999}. Although both modes are observed 
at a slightly higher energy than the `LSDA + model' value (table 1), the observed 
9\% anisotropy is within the suggested bounds of 3-10\% \cite{massiddaetal1999}.

Plots of the temperature evolution of the ZC phonon anisotropy 
are shown in figures \ref{zcsplitting}(inset) and \ref{fulltdep}. 
The splitting in the phonon spectra is clearly associated with the transition 
to magnetic order. Asymmetry in peaks 
measured at 130 K and 150 K (fig. \ref{fulltdep}) hints that a small amount of anisotropy 
may be present at temperatures substantially above $\mathrm{T_{N}}$, (i.e. when 
the unit cell is thought to be perfectly cubic \cite{Morosin1970}). 

\begin{figure}
\begin{center}
\resizebox{75mm}{!}{\includegraphics{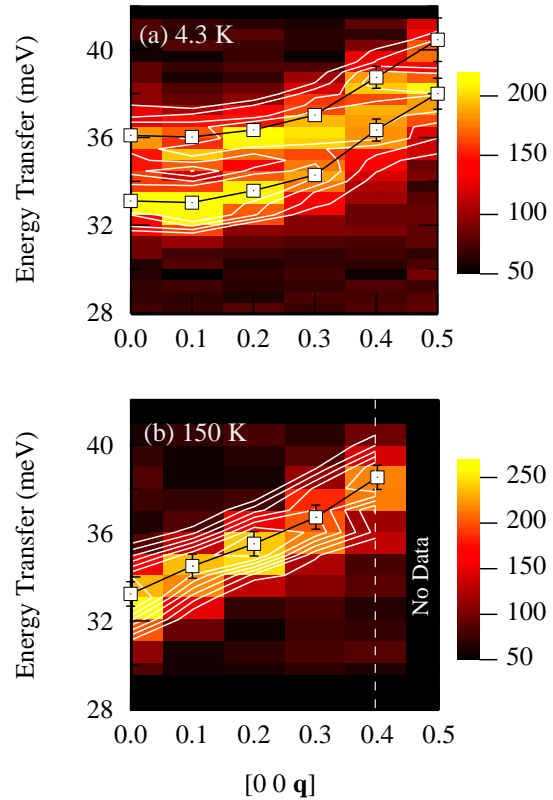}}
\caption{Dispersion curves for TO phonon modes measured at (a) 4.3 K and (b) 150 K along the [0 0 1] direction. 
The $\mathbf{q}=0$ position in both graphs corresponds to the (2 2 0) ZC. 
Markers indicate the peak positions of Gaussian fits to the data. }
\label{disp1}
\end{center}
\end{figure}

\begin{figure}
\begin{center}
\resizebox{8cm}{!}{\includegraphics{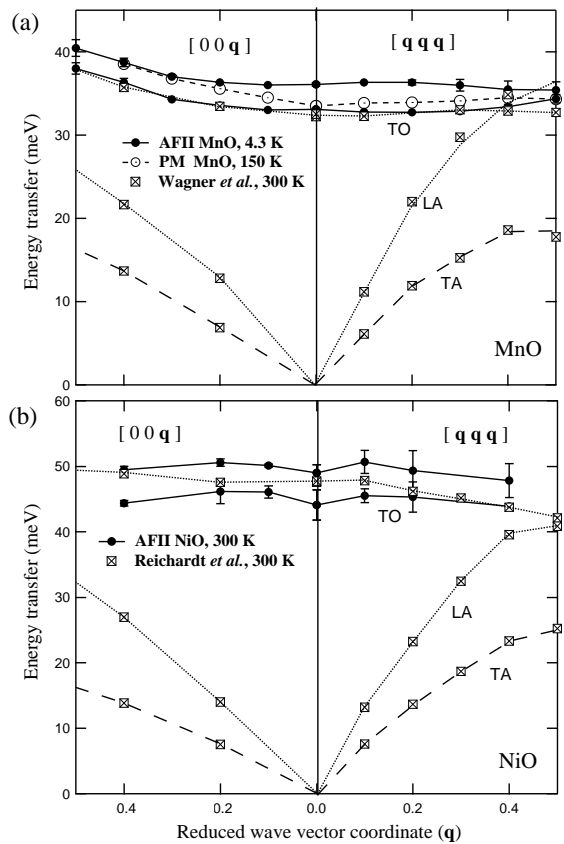}}
\caption{(a) MnO phonon dispersion curves measured at 4.3 K (filled circles) and 150 K 
(open circles). Square markers show the room temperature data of Wagner \emph{et al.}
\cite{Wagneretal1976}. (b) Room temperature TO phonon dispersion curves measured for 
NiO. Filled round markers show measurements along the [001] and [111] directions, 
(obtained around \textbf{Q}=(220) and \textbf{Q}=(222)). Square markers show the 
data of Reichardt \emph{et al.} \cite{Reichardtetal1975}.}
\label{disp2}
\end{center}
\end{figure}

Dispersion curves obtained for TO modes along the [100] and [111] directions in MnO at 4.3 K and 150 K are presented in figures \ref{disp1} and \ref{disp2}(a). 
Although the splitting extends throughout the Brillouin zone, the distortion is most 
clearly resolved at values of reduced wavevector (\textbf{q}) close to the zone-center, see figure \ref{disp1}(a).
Further from the ZC (where the scattering is less intense and the spectrometer 
may be less focussed), the presence of two modes is revealed by the deviation 
of the shape of the overall peak from Gaussian, and its 
increased width in comparison to the same measurement at 150 K, fig. \ref{disp1}(b). As the crystal consisted of two domains, intensity from differently polarized magnons could be seen reasonably well in all of the measured directions.  

The notion that a 
small splitting may be present in MnO above T$_{N}$ is supported by the dispersion 
measured at 150 K, shown in figure 4(a). The 150 K data (open circles) 
was fitted to a single Gaussian peak but found to be slightly raised in energy in comparison 
to the lower energy mode (filled circles), and to the room temperature data of 
Wagner \emph{et al.} \cite{Wagneretal1976} (square markers). 
 
To investigate whether a similar phonon splitting could be observed in the room temperature 
dispersion of NiO, a preliminary survey was performed using the HB1 spectrometer located 
at the High Flux Isotope Reactor, Oak Ridge National Laboratory, USA. As before, 
the crystal was oriented with the [110] direction vertical, and the spectrometer 
operated in constant-\emph{Q} mode with fixed final energy $\mathrm{E_{f}}$ = 14.73 meV.  

Our NiO crystal was of less reliable quality than the MnO crystal, and spurious 
`Bragg-Bragg-inelastic' scattering was observed. This limited our measurements to 
regions surrounding ZC's located at \textbf{Q}=(220) and \textbf{Q}=(222). 
Room temperature TO dispersion curves for the [111] and [001] directions of 
antiferromagnetically ordered NiO are presented in figure 4(b)(filled round markers). Squares represent 
the data of Reichardt \emph{et al.} \cite{Reichardtetal1975}, also collected at room temperature.

In our NiO data two peaks were observed occurring at energies of around 45 and 50 meV. 
This is in reasonable agreement with previous reports indicating a single TO excitation 
with an energy varying between 46 and 50 meV \cite{Coyetal1976,Reichardtetal1975}. 
This ZC splitting of approximately 5 meV is larger in absolute magnitude than observed in MnO, (despite the smaller distortion angle), 
and corresponds to roughly 10\% of the average energy of the two modes. 

Since both this data and data of previous researchers were collected 
at room temperature (where NiO is already magnetically ordered) it is necessary to 
ask why this effect had not been been observed before. We suggest that the cause of the 
discrepancy lies in the poor energy resolution and flux of the early spectrometers, 
particularly at high energy transfers. These experimental factors coupled with the expectation of a single (degenerate) phonon mode may have lead to fitting 
of the data to a single peak. 

In summary, we have shown that below $\mathrm{T_{N}}$ the dynamic properties of MnO and 
NiO are substantially non-cubic. The magnitude of the zone center 
anisotropy observed in transverse optic modes in MnO and NiO corresponds to 
9-10\% of the energy of the original (degenerate) phonon. 
The intensities of the two optic modes in MnO and the form of the dispersion are consistent 
with the picture of the splitting described by recent band structure calculations where a 
higher energy mode is predicted for phonons polarized along [111] \cite{massiddaetal1999}. 

In MnO, we have shown that the splitting of the phonon modes is associated 
with the transition to magnetic order. 
Although exchange-striction induces a small rhombohedral distortion, this static 
distortion in itself is too weak to explain the magnitude of the anisotropy. 
Furthermore, NiO possesses a significantly smaller distortion angle than MnO but a larger (5 meV) splitting. 
The strain associated with
exchange-striction (even in the undistorted 
structure considered by Massidda \emph{et al}. \cite{massiddaetal1999}) defines a 
new potential well for the atoms, which may significantly influence the dynamics.    
However, further research will be required to investigate the mechanism underlying the splitting and  
determine how the magnetic-order induced phonon anisotropy in MnO and NiO is mediated. 
Theoretical studies investigating 
the r\^{o}le of electronic correlations on the phonon modes 
of NiO are particularly encouraged. 
Although theorists have not provided any calculations for 
NiO as yet, we are confident that our preliminary findings of an even larger splitting will provide 
a new impetus for them to do so. 

We thank  Sandro Massidda, Julie Staunton, Jim Hague, Randy Fishman and Oleg 
Petrenko for useful discussions, 
and the Engineering and Physical Sciences Research Council, UK, for financial support. 

\bibliography{mno}

\end{document}